\documentclass[aps,prd,reprint,showpacs,superscriptaddress]{revtex4-2}

\usepackage{amsmath}
\usepackage{amssymb}
\usepackage{graphicx}
\usepackage{hyperref}
\usepackage{xcolor}

\begin{document}

\title{Matter Creation in Unimodular Gravity Cosmology}

\author{V\'ictor H. C\'ardenas}
\affiliation{Instituto de F\'isica y Astronom\'ia, Facultad de Ciencias, Universidad de Valpara\'iso, Gran Breta\~na 1111, Valpara\'iso, Chile}
\email{victor.cardenas@uv.cl}

\author{Miguel Cruz}
\affiliation{Facultad de F\'isica, Universidad Veracruzana, 91097 Xalapa, Veracruz, M\'exico}
\email{miguelcruz02@uv.mx}

\author{Samuel Lepe}
\affiliation{Instituto de F\'isica, Pontificia Universidad Cat\'olica de Valpara\'iso, Avda. Brasil 2950, Valpara\'iso, Chile}
\email{samuel.lepe@pucv.cl}

\date{\today}

\begin{abstract}
We investigate matter creation in the cosmological framework of unimodular gravity. We keep
the energy density $\rho$, pressure $p$, particle number density $n$, and particle number
$N=nV$ as the physical thermodynamic variables of the matter sector, and introduce particle
creation directly through the balance equation $N^{\mu}{}_{;\mu}=n\Gamma$. Combining this
open-system formulation with the unimodular diffusion term $Q$, we derive the exact identity
$nT\dot\sigma=-3Hp_c-(\rho+p)\Gamma-\dot Q$, which shows that the creation pressure $p_c$, the creation rate $\Gamma$, and the diffusion term $Q$ are, in general, three independent sources of entropy production per particle. Adopting the standard postulate that particle creation is itself adiabatic, the first two terms cancel identically and the relation reduces to the particular case $nT\dot\sigma=-\dot Q$, so that the entropy per particle is constant only when $Q$ is constant. Thus, even when the creation process associated with $\Gamma$ is adiabatic in the usual Prigogine sense, the full matter evolution in unimodular gravity is non-adiabatic whenever $\dot Q\neq0$. The general identity, however, remains available as a starting point for relaxing this postulate in future work.
\end{abstract}

\maketitle

\section{Introduction}

The standard cosmological model provides a remarkably successful description of the
large-scale evolution of the Universe. Nevertheless, the physical origin of the observed
accelerated expansion remains one of the central open problems in modern cosmology~\cite{Riess1998,Perlmutter1999}. The
simplest explanation is a cosmological constant, but this interpretation faces conceptual
difficulties related to its magnitude, its radiative stability, and its interpretation as
vacuum energy~\cite{Weinberg1989}.

Unimodular gravity offers an alternative perspective on the cosmological constant problem.
In this framework, the determinant of the metric is fixed, and the cosmological constant
appears as an integration constant rather than as a fundamental parameter in the action~\cite{vanderBij1982,Buchmuller1988,HenneauxTeitelboim1989,Unruh1989,Ellis2011, Einstein1919, Ellis2014, NgVanDam1991, Kuchar1991, Smolin2009, Alvarez2005, CarballoRubio2022}.
At the level of cosmology, unimodular gravity may also lead to modified conservation
equations in which the matter sector exchanges energy with an additional contribution
usually encoded in a diffusion term $Q$.

The purpose of this work is to study matter creation in unimodular gravity cosmology.
Matter creation in an expanding universe can be formulated using the thermodynamics of
open systems. In this approach, the particle number is not conserved, and the particle
creation rate $\Gamma$ induces an effective negative pressure, usually called the creation
pressure. This mechanism has been widely discussed in the context of gravitational particle
production and cosmological models with irreversible matter creation~\cite{ref2,ref3,LimaGermano1992, Parker1969, Hu1982, LimaGermanoAbramo1996, LimaBaranov2014, Steigman2009, NunesPavon2015, Schiavone2026}.

In unimodular gravity, however, the situation is more subtle. The matter sector is already
non-conserved due to the unimodular diffusion term $Q$. While Ref.~\cite{us} developed the thermodynamics of unimodular gravity with the diffusion term $Q$, and Ref.~\cite{ref4} studied the thermodynamics of matter creation including adiabaticity and phantom behavior, the present work synthesizes these two ingredients. Therefore, one must distinguish
between two effects: particle creation, controlled by $\Gamma$, and unimodular diffusion,
controlled by $Q$. A central point of this paper is that these two effects have different
physical meanings. The creation rate $\Gamma$ modifies the effective pressure of the matter
sector, whereas $Q$ acts as an independent source of non-adiabaticity.

We shall focus on the physical matter interpretation. That is, we do not absorb $Q$ into an
effective conserved fluid, nor do we reinterpret it as an independent vacuum-energy
component. Instead, we keep $\rho$, $p$, $n$, and $N=nV$ (with $V$ being the comoving volume) as the physical thermodynamic
variables of the matter sector, and we introduce particle creation directly through the
balance equation for the particle number current $N^\mu = n u^\mu$:
\begin{equation}
N^{\mu}{}_{;\mu} = n\Gamma. \label{eq:1}
\end{equation}
Within this route, the central structural result of this work is the exact identity
\begin{equation}
nT\dot\sigma = -3Hp_c - (\rho+p)\Gamma - \dot Q, \label{eq:2}
\end{equation}
which shows that the creation pressure $p_c$, the creation rate $\Gamma$, and the diffusion
term $Q$ are, in general, three independent sources of entropy production per particle. The
commonly used relation $nT\dot\sigma=-\dot Q$ is recovered only as a particular closure of
Eq.~\eqref{eq:2} --- obtained by postulating that particle creation is itself adiabatic ---
and is one among several physically motivated closures examined below. That particular
closure already illustrates the main qualitative point: the entropy per particle is
constant only when $Q$ is constant, so that even when particle creation itself is adiabatic
in the usual sense, the full matter evolution in unimodular gravity is generically
non-adiabatic whenever $\dot Q\neq0$.

The paper is organized as follows. In Sec.~\ref{sec:UG} we review the basic cosmological
equations of unimodular gravity. In Sec.~\ref{sec:thermo} we discuss the thermodynamic role
of the diffusion term $Q$. In Sec.~\ref{sec:creation} we introduce particle creation in the
physical matter sector and derive the creation pressure. In Sec.~\ref{sec:combined} we
combine matter creation with unimodular diffusion, derive the general entropy-production
identity, and examine five particular closures of that identity, ranging from the case in
which $Q$ remains fully independent of $\Gamma$ to a causal, relaxation-based closure for
$p_c$, together with a reconstruction of $\Gamma$ from a prescribed dark-energy background. Additionally, to explicitly demonstrate that the separation between matter creation and unimodular diffusion is operationally meaningful, we integrate the background dynamics for a causally evolving creation pressure using an observationally motivated diffusion model. Sec.~\ref{sec:discussion} compares these closures and discusses their physical
content, and Sec.~\ref{sec:conclusions} contains our conclusions.

\section{Unimodular Cosmology}
\label{sec:UG}

We consider a spatially flat FLRW spacetime,
\begin{equation}
ds^2 = -dt^2 + a^2(t)\,d\mathbf{x}^2, \label{eq:3}
\end{equation}
where $a(t)$ is the scale factor and $H=\dot a/a$ is the Hubble parameter. While the standard FLRW metric in cosmic time yields $\sqrt{-g} = a^3(t)$, the fundamental unimodular condition $\sqrt{-g} = 1$ can be strictly satisfied by adopting a unimodular time coordinate $d\tau = a^3 dt$. Alternatively, in fully covariant formulations, this is handled via a Lagrange multiplier coupled to a fixed fiducial volume element. In either approach, the resulting background Friedmann equations, when expressed in terms of the physical cosmic time $t$, remain exactly structurally unaffected \cite{Smolin2009,Ellis2014}. In units
$8\pi G=c=1$, and setting the integration cosmological constant to zero for simplicity, the
background equations of unimodular gravity can be written as \cite{Josset:2016vrq,DeAngelis2026,Calogero2011,GarciaAspeitia2019,Casas2018,GarciaBellido2011}
\begin{align}
3H^2 &= \rho + Q, \label{eq:4}\\
2\dot H + 3H^2 &= -p + Q, \label{eq:5}\\
\dot\rho + 3H(\rho+p) &= -\dot Q. \label{eq:6}
\end{align}
Here $\rho$ and $p$ denote the energy density and pressure of the physical matter sector,
while $Q$ is the unimodular diffusion term.

The combination of Eqs.~\eqref{eq:4} and \eqref{eq:5} gives
\begin{equation}
\dot H = -\tfrac{1}{2}(\rho+p). \label{eq:7}
\end{equation}
Thus, the Raychaudhuri equation keeps the same form as in standard general relativity,
while the Friedmann equation receives an additional contribution through $Q$. Equation
\eqref{eq:6} shows that the physical matter sector is not conserved whenever $\dot Q\neq0$.
Equivalently, the total effective fluid defined by $\rho_{\rm eff} = \rho+Q$ and $p_{\rm eff} = p-Q$ is conserved. However, in this work we shall not use this effective-fluid picture as the fundamental interpretation. Instead, we shall keep $\rho$ and $p$ as the physical matter variables, and we shall treat $Q$ as a diffusion term modifying the balance equation of the matter sector.

It is useful to introduce the deceleration parameter $q \equiv -1 - \frac{\dot H}{H^2}$. Using Eqs.~\eqref{eq:4} and \eqref{eq:7}, for a barotropic fluid $p = \omega\rho$, one obtains
\begin{equation}
q = -1 + \frac{3}{2}(1+\omega)\,\frac{\rho}{\rho+Q}. \label{eq:13}
\end{equation}
Therefore, even before introducing particle creation, the diffusion term $Q$ modifies the
background expansion by contributing to the total energy density that sources the
Friedmann equation.

\section{Thermodynamic Role of the Unimodular Diffusion Term}
\label{sec:thermo}

Let us consider a comoving volume $V=V_0 a^3$. For a closed matter sector with internal
energy $U$ the first law is $T\,dS = dU + p\,dV$. Using $U=\rho V$, one obtains
\begin{equation}
T\,dS = V\left[d\rho + 3H(\rho+p)\,dt\right]. \label{eq:15}
\end{equation}
Then, using the unimodular balance equation \eqref{eq:6}, one finds
\begin{equation}
T\,dS = -V\,dQ. \label{eq:16}
\end{equation}
Thus, the diffusion term $Q$ controls the entropy production of the matter sector. In
particular,
\begin{equation}
\dot S = -\frac{V}{T}\dot Q. \label{eq:17}
\end{equation}
For positive temperature, as examined in \cite{us}, fulfillment of the second law demands that
\begin{equation}
\dot S \geq 0 \;\Rightarrow\; \dot Q \leq 0. \label{eq:18}
\end{equation}
Equivalently, in terms of redshift, the condition $\dot Q\leq0$ becomes
\begin{equation}
\frac{dQ}{dz} \geq 0. \label{eq:20}
\end{equation}
Equation \eqref{eq:16} is important because it shows that, in the physical matter
interpretation, unimodular cosmology is generically non-adiabatic. The case $\dot Q=0$
corresponds to a constant contribution in the Friedmann equation, which is indistinguishable
from a cosmological constant at the background level. The genuinely unimodular-diffusive
case corresponds to $\dot Q\neq0$.

\section{Particle Creation in the Physical Matter Sector}
\label{sec:creation}

We now introduce particle creation directly in the physical matter sector. Particle creation is described by Eq.~\eqref{eq:1}. In a FLRW spacetime, this becomes
\begin{equation}
\dot n + 3Hn = n\Gamma. \label{eq:23}
\end{equation}
The thermodynamics of an open system is governed by
\begin{equation}
T\,dS = dU + p\,dV - \mu\,dN, \label{eq:26}
\end{equation}
where $\mu$ is the chemical potential. Introducing the densities
$\rho = \frac{U}{V}$, $n = \frac{N}{V}$, and $s = \frac{S}{V} = n\sigma$,
where $\sigma=S/N$ is the entropy per particle, Eq.~\eqref{eq:26} leads to $T\,ds = d\rho - \mu\,dn$.
Using the Euler relation $\rho+p = Ts + \mu n$, one obtains the Gibbs identity
\begin{equation}
nT\,d\sigma = d\rho - \frac{\rho+p}{n}\,dn, \label{eq:30}
\end{equation}
a relation that remains valid, as shown below, even in the presence of unimodular diffusion.

In the standard formulation of adiabatic particle creation in general relativity, one assumes that the entropy per particle is constant: $\dot\sigma = 0$. Using Eq.~\eqref{eq:30} and Eq.~\eqref{eq:23}, this can be rewritten as
\begin{equation}
\dot\rho + 3H(\rho+p+p_c) = 0, \label{eq:34}
\end{equation}
where
\begin{equation}
p_c = -(\rho+p)\,\frac{\Gamma}{3H} \label{eq:35}
\end{equation}
is the creation pressure. For a barotropic fluid $p=\omega\rho$, this becomes
$p_c = -(1+\omega)\rho\,\frac{\Gamma}{3H}$.
Thus, for $\Gamma>0$ and $\rho+p>0$, the creation pressure is negative. This is the usual
mechanism by which gravitationally induced particle creation can contribute to the
accelerated expansion.

\section{Matter Creation in Unimodular Gravity}
\label{sec:combined}

We now combine the previous open-system formulation with the unimodular balance equation. The natural generalization of the matter balance equation, keeping $p_c$ general for the
moment, is
\begin{equation}
\dot\rho + 3H(\rho+p+p_c) = -\dot Q. \label{eq:37}
\end{equation}
Substituting Eqs.~\eqref{eq:37} and \eqref{eq:23} into the exact Gibbs identity
\eqref{eq:30}, without yet specifying $p_c$, we recover the exact identity stated in Eq.~\eqref{eq:2}:
\begin{equation}
nT\dot\sigma = -3Hp_c - (\rho+p)\Gamma - \dot Q. \label{eq:40gen}
\end{equation}
This exact identity shows that there are, in general, three independent contributions to the
production of entropy per particle: the creation pressure $p_c$, the creation rate $\Gamma$, and the unimodular diffusion term $Q$. Note that a complete integration of the entropy per particle $\sigma(z)$ requires specifying an additional thermodynamic equation of state for the temperature $T(n, \rho)$.

From Eq.~\eqref{eq:40gen}, different physical closures correspond to different ways of distributing the three source terms. We examine five particular cases: the first four restrict throughout to a barotropic fluid $p=\omega\rho$ and the standard creation-rate Ansatz $\Gamma=3\beta H$, with $\beta$ a constant ($\beta \ge 0$, with $\beta>1$ corresponding to the phantom regime), while the fifth reconstructs $\Gamma$ from a prescribed dark-energy background.

\subsection{Case (0): $\Gamma$-only creation pressure, $Q$ independent}
\label{sec:case0}

This is the closure referred to as the ``standard postulate'' in the abstract: particle creation, by itself, does not generate entropy per particle, so that $p_c$ is fixed by $\Gamma$ alone through Eq.~\eqref{eq:35}. Here $Q$ is \emph{not} required to satisfy any relation involving $\Gamma$; it remains an independent background function, subject only to the second-law condition of Sec.~\ref{sec:thermo}. This is the case for which we now work out the resulting background dynamics.

The full background system is governed by Eqs.~\eqref{eq:4}, \eqref{eq:5}, and \eqref{eq:37}, with $p_c$ given by Eq.~\eqref{eq:35}. The Raychaudhuri equation becomes
\begin{equation}
\dot H = -\tfrac{1}{2}(\rho+p+p_c). \label{eq:d2}
\end{equation}
Therefore, the deceleration parameter is
\begin{equation}
q = -1 + \frac{3}{2}(1+\omega)\,\frac{\rho}{\rho+Q}\left(1-\frac{\Gamma}{3H}\right). \label{eq:d5}
\end{equation}
This expression displays the two distinct mechanisms affecting cosmic acceleration in this
closure: $Q$ contributes to the effective energy density sourcing the Friedmann equation,
while $\Gamma$ reduces the effective enthalpy of the physical matter sector. The effective equation-of-state parameter $\omega_{\rm eff}$ is given by
\begin{equation}
\omega_{\rm eff} = -1 + (1+\omega)\,\frac{\rho}{\rho+Q}\left(1-\frac{\Gamma}{3H}\right). \label{eq:d9}
\end{equation}
Assuming $H>0$, $\rho>0$, $\omega>-1$, and $\rho+Q>0$, the condition $\Gamma > 3H$ is in fact both necessary and sufficient for $\omega_{\rm eff} < -1$. Such a regime corresponds to a very strong particle production rate and should be treated with care \cite{ref4}.

In terms of redshift, the matter balance equation becomes
\begin{equation}
\frac{d\rho}{dz} - \frac{3(1+\omega)}{1+z}\left(1-\frac{\Gamma}{3H}\right)\rho = -\frac{dQ}{dz}, \label{eq:d13}
\end{equation}
which makes explicit that $Q(z)$ and $\Gamma(z)$ cannot be chosen independently of each
other if a thermodynamically consistent evolution is required, since the second-law
condition $dQ/dz\geq0$, Eq.~\eqref{eq:20}, remains active in this closure. The entropy per particle satisfies, in redshift,

\begin{equation}
nT\frac{d\sigma}{dz} = -\frac{dQ}{dz}, \label{eq:d16}
\end{equation}
consistent with $\sigma$ decreasing with redshift (and therefore increasing with cosmic time) whenever $dQ/dz\geq0$.

\subsection{Case (i): global adiabaticity, $\dot\sigma=0$}

Setting $\dot\sigma=0$ in Eq.~\eqref{eq:40gen} fixes $p_c$ as a function of \emph{both} $\Gamma$ and $Q$:
\begin{equation}
p_c = -\frac{(\rho+p)\Gamma+\dot Q}{3H}. \label{eq:pcgen}
\end{equation}
Substituting this into the general balance equation~\eqref{eq:37}, the $\dot Q$ terms cancel \emph{identically}, leaving
\begin{equation}\label{eq:c3}
\dot\rho + 3H(\rho+p) = (\rho+p)\Gamma.
\end{equation}
This is a nontrivial result: under global adiabaticity, the physical energy density $\rho$
obeys exactly the ordinary Prigogine creation equation, \emph{completely independent of
$Q$}.

With $p=\omega\rho$ and $\Gamma=3\beta H$, Eq.~(\ref{eq:c3}) integrates to $\rho(z) = \rho_0(1+z)^{3(1+\omega)(1-\beta)}$. Since Friedmann's equation is untouched, $H(z)$ follows directly once $Q(z)$ is specified:
\begin{equation}\label{eq:c5}
3H^2(z) = \rho_0(1+z)^{3(1+\omega)(1-\beta)} + Q(z).\,
\end{equation}
For $\omega=0$ and constant $Q$, this reproduces the familiar CCDM-type expansion history~\cite{LimaJesusOliveira2010}.

\emph{Remark on the second law.} With $\dot\sigma=0$, the time variation of entropy gives $\dot S=N\sigma\Gamma$: all entropy production is sourced by $\Gamma$ alone, and the second law $\dot S\geq0$ only requires $\Gamma\geq0$ (assuming $\sigma \ge 0$). In this closure the sign condition $\dot Q\leq0$ is \emph{not} required --- $Q(z)$ is thermodynamically unconstrained.

\subsection{Case (ii): no creation pressure, $p_c\equiv0$}

Here we postulate that particle creation does not source any effective pressure at all: the
background dynamics is left exactly as in Sec.~\ref{sec:UG}--\ref{sec:thermo}, unmodified by
$\Gamma$, while $\Gamma$ remains thermodynamically active only through Eq.~\eqref{eq:40gen}, which becomes
\begin{equation}\label{eq:c7}
nT\dot\sigma = -(\rho+p)\Gamma-\dot Q.
\end{equation}
Since $p_c=0$, the balance equation reduces to the \emph{pure diffusion}
equation $\dot\rho+3H(\rho+p)=-\dot Q$, which is completely insensitive to $\Gamma$.
In redshift, for $p=\omega\rho$, the general solution leads to
\begin{equation}\label{eq:c11}
3H^2(z) = (1+z)^3\left[\rho_0-\int_0^z(1+z')^{-3}\frac{dQ}{dz'}dz'\right] + Q(z). 
\end{equation}
This closure is intended primarily as a formal limiting case for $\omega=0$. Creating particles with non-zero rest mass requires energy; with $p_c=0$, this energy must be supplied entirely by the diffusion term $Q$. Consequently, for a thermodynamically consistent evolution ($\dot{S} \ge 0$), the energy inflow from diffusion ($\dot{Q} < 0$) must be sufficient to cover the cost of creation. When $Q$ is constant, however, there is no energy inflow, leading to an unphysical unbounded decrease in energy per particle. To see this explicitly, consider the evolution of the energy per particle $\epsilon \equiv \rho/n$. Using the balance equations $\dot{\rho} + 3H\rho = -\dot{Q}$ (for $p=p_c=0$) and $\dot{n} + 3Hn = n\Gamma$, the time derivative of $\epsilon$ yields $\dot{\epsilon} = -\dot{Q}/n - \Gamma\epsilon$. If the diffusion term is constant ($\dot{Q} = 0$), the energy per particle decays exponentially, $\dot{\epsilon} = -\Gamma\epsilon$ for $\Gamma>0$. For massive matter, $\epsilon$ is strictly bounded from below by the rest mass. An unbounded decrease is therefore unphysical, demonstrating that sustaining matter creation in this closure strictly requires a continuous energy inflow from the unimodular diffusion term ($\dot{Q} < 0$) to provide the rest-mass energy of the newly created particles. For a radiation-like fluid, the rest-mass argument does not apply, and the case reduces harmlessly. In this closure, $\Gamma$ is thermodynamically active but \emph{dynamically inert}.

\subsection{Case (iii): causal relaxation for $p_c$}

Rather than an algebraic closure, $p_c$ is now promoted to an independent dynamical variable
obeying a truncated causal transport equation (the same structure used in the
bulk-viscosity description of particle creation, e.g.,~\cite{Zimdahl:1996,Maartens:1995,IsraelStewart1979}),
\begin{equation}\label{eq:c12}
\tau\dot p_c + p_c = -3H\xi, 
\end{equation}
where $\tau$ is a relaxation time and $\xi$ an effective viscosity coefficient. The general identity~\eqref{eq:40gen} remains exact and unmodified; what changes is that $p_c$ is no longer slaved algebraically to $\Gamma$ and $Q$. In the instantaneous ($\tau \to 0$) limit, the system plainly recovers Case (0). The full system in redshift variables reads
\begin{equation}
3H^2 = \rho+Q(z), \label{eq:c14}
\end{equation}
\begin{equation}\label{eq:c15}
\frac{d\rho}{dz} = \frac{3(1+\omega)\rho}{1+z} + \frac{3p_c}{1+z} - \frac{dQ}{dz}, 
\end{equation}
\begin{equation}\label{eq:c16}
\frac{dp_c}{dz} = \frac{3H\xi+p_c}{\tau(1+z)H}, \qquad \xi=\frac{\beta(1+\omega)\rho}{3H}.
\end{equation}
Because $p_c$ is now evolving independently, the second law $\dot{S} \ge 0$ imposes a nontrivial thermodynamic constraint on $p_c, \Gamma,$ and $Q$ through Eq.~\eqref{eq:40gen}. In particular, in the $\tau \to \infty$ limit where $p_c$ freezes, the evolution must still satisfy $-3Hp_c - (\rho+p)\Gamma - \dot{Q} \ge 0$ to be thermodynamically admissible.

\subsection{Case (iv): reconstruction of $\Gamma$ from a dark-energy background}
\label{sec:emulation}

In the four closures examined above, the creation rate was fixed through the phenomenological Ansatz $\Gamma=3\beta H$. Here we present an alternative in which $\Gamma$ is \emph{reconstructed} by requiring that the model reproduce the background expansion of a prescribed dark energy (DE) model. This is the unimodular realization of the background degeneracy between matter creation and DE established, within general relativity, in Ref.~\cite{CardenasCruz2024}.

The Friedmann constraint fixes the sum of the physical matter density $\rho$ and the diffusion term $Q$ to $3H^2 = \rho_m+\rho_x$, with $p_x=w(a)\,\rho_x$ and $\rho_m=\rho_{m0}\,a^{-3}$.
A constant diffusion, $\dot Q = 0$, cannot be distinguished from a cosmological constant at the background level. This suggests the physically motivated split
\begin{equation}
\rho_x=\rho_\Lambda+\rho_d(a),\qquad
Q=\rho_\Lambda\equiv\Lambda=\text{const},
\label{eq:emu-split}
\end{equation}
in which the constant ($w=-1$) piece of the dark sector is supplied geometrically by the unimodular constant $\Lambda$, while the genuinely dynamical part $\rho_d(a)$ is generated by particle creation. For pressureless created matter ($p=0$) with constant diffusion, we find
\begin{equation}
\Gamma(a)=-\,\frac{3H\,w_d(a)\,\rho_d(a)}{\rho_m(a)+\rho_d(a)}.
\label{eq:emu-Gamma}
\end{equation}
Since the second law requires $\Gamma \ge 0$ (assuming $\sigma \ge 0$), Eq.~\eqref{eq:emu-Gamma} demands that the product $w_d(a)\rho_d(a)$ must be non-positive. Assuming on standard physical grounds that the dynamical dark energy density is strictly positive ($\rho_d > 0$), the sign of the creation rate is determined entirely by the equation-of-state parameter $w_d(a)$, which must therefore be non-positive.

For a pure cosmological constant, $\rho_d=0$, Eq.~\eqref{eq:emu-Gamma} gives $\Gamma=0$: $\Lambda$CDM is reproduced with \emph{vanishing} matter creation. At the background level, setting $Q=\Lambda=\text{const}$ is mathematically equivalent to general relativity with a bare cosmological constant. This provides a conceptual distinction; however, in unimodular gravity, $\Lambda$ arises unavoidably as an integration constant rather than being inserted by choice, separating the purely geometric dark sector from the dynamical creation sector.

\section{Discussion}
\label{sec:discussion}

The general identity~\eqref{eq:40gen} shows that the creation pressure $p_c$, the creation
rate $\Gamma$, and the unimodular diffusion term $Q$ are, in principle, three independent
sources of entropy production per particle. 

In Case~(0), $\Gamma$ fixes $p_c$ on its own, exactly as in ordinary Prigogine-type
creation \cite{ref2}, while $Q$ remains an independent background function. In Case~(i), imposing global adiabaticity $\dot\sigma=0$ redistributes the entropy budget so that $\Gamma$ alone guarantees $\dot S\geq0$, completely decoupling the matter density from $Q$. In Case~(ii), a vanishing creation pressure renders $\Gamma$ dynamically inert while leaving it thermodynamically active. Case~(iii) relaxes the algebraic slaving of $p_c$ to $\Gamma$ altogether, promoting it to an independently evolving quantity. Finally, Case~(iv) abandons the Ansatz altogether: rather than prescribing $\Gamma$, it reconstructs the creation rate by requiring the model to reproduce a target dark energy background.

Particle creation of the Prigogine--Calv\~ao--Lima--Waga type leaves the physical sector energy-momentum tensor conserved, and therefore cannot source the unimodular diffusion current that defines $Q$ through the geometric identity underlying unimodular gravity~\cite{Josset:2016vrq}. Consequently, $Q(z)$ must continue to be specified phenomenologically. In this respect, the anomalous-decay and sudden-transfer forms of $Q(z)$ proposed to address the $H_0$ tension~\cite{Perez:2020cwa} provide observationally motivated candidates.

\begin{center}
\begin{figure}[htbp!]
\includegraphics[scale=1]{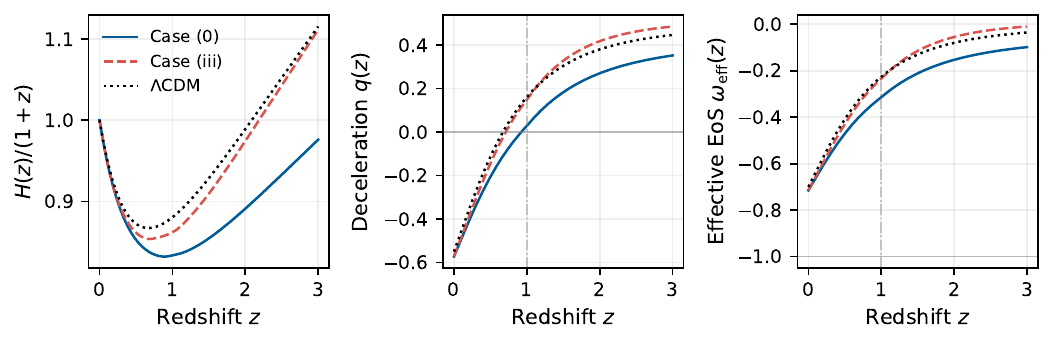}
\caption{Background evolution for Case (0) and Case (iii) compared to $\Lambda$CDM. We use the anomalous decay model for $Q(z)$~\cite{Perez:2020cwa} with a transition redshift $z^\star = 1.0$. Both closures break the background degeneracy for $z > z^\star$, with the causal closure (Case iii) showing a distinct relaxation behavior in the deceleration parameter.}
\label{fig:dynamics}
\end{figure}
\end{center}

To explicitly illustrate that the $\Gamma-Q$ separation is operationally meaningful and that the background degeneracy can be broken, we can numerically integrate the system using an observationally motivated form for $Q(z)$. Following the phenomenological ``anomalous decay'' model proposed in \cite{Perez:2020cwa} to alleviate the $H_0$ tension, we assume that the matter sector cedes energy to the dark sector at late times, starting at a transition redshift $z^{\star} = 1.0$. Setting $\Gamma=3\beta H$ and using the exact piecewise diffusion rate $dQ/dz$ from Ref.~\cite{Perez:2020cwa}, we integrate the background evolution for both the algebraic closure (Case 0) and the causal relaxation closure (Case iii). As shown in Fig.~\ref{fig:dynamics}, while a strictly constant $Q$ mimics a cosmological constant, introducing a dynamically evolving $Q(z)$ breaks the background degeneracy. The numerical integration of $H(z)$, the deceleration parameter $q(z)$, and the effective equation of state $\omega_{\text{eff}}(z)$ demonstrates that both closures deviate distinctively from $\Lambda$CDM at $z > z^\star$, with the causal closure exhibiting a characteristic relaxation delay in $q(z)$ governed by $\tau$. Thus, $\Gamma$ and $Q$ imprint distinguishable signatures on the expansion history when dynamically active.
\onecolumngrid
\twocolumngrid

\section{Conclusions}
\label{sec:conclusions}

We have studied matter creation in unimodular gravity cosmology by introducing particle
production directly in the physical matter sector. Combining the open-system formulation of
particle creation with the unimodular diffusion term $Q$, we derived the exact identity
\begin{equation}
nT\dot\sigma = -3Hp_c - (\rho+p)\Gamma - \dot Q, \label{eq:concl1}
\end{equation}
which shows that $p_c$, $\Gamma$, and $Q$ are generally three independent sources of
entropy production per particle. We examined five physically motivated closures of Eq.~\eqref{eq:concl1}, finding that $\Gamma$ and $Q$ should generically be regarded as independent physical channels, rather than as two phenomenological descriptions of a single underlying mechanism.

Ultimately, our results establish that matter creation and unimodular diffusion must be treated as fundamentally distinct thermodynamic processes. Because particle production in the standard open-system formulation leaves the physical energy-momentum tensor conserved, it cannot act as the physical source of the diffusion term $Q$, which originates purely from the geometric constraints of unimodular gravity. Recognizing $\Gamma$ and $Q$ as independent evolutionary channels provides a rigorous thermodynamic foundation for future theoretical explorations of the dark sector, ensuring that geometric modifications to gravity are not confused with the irreversible thermodynamics of matter creation.

\begin{acknowledgments}
V.H.C. acknowledges support from CEFITEV-UV. M.C. acknowledges partial support from
S.N.I.I. (SECIHTI-M\'exico). S.L. acknowledges support from FONDECYT grant No.~1250969 from
the Government of Chile.
\end{acknowledgments}

\end{document}